\def\'#1{\ifx#1i{\accent"13\i}\else{\accent"13#1}\fi}

\def\be{\begin{equation}}

\def\ee{\end{equation}}

\def\lsd{LSD}

\def\ssd{SSD}
\documentclass[12pt,preprint]{aastex}

\usepackage{graphicx}
\usepackage{ae}

\usepackage{natbib}

\begin{document}

\title{Quiescent and coherent cores from
  gravoturbulent fragmentation}

\author{Ralf S.\ Klessen$^{1}$, Javier Ballesteros-Paredes$^{2}$, 
Enrique V\'azquez-Semadeni$^2$, and Carolina Dur\'an-Rojas$^2$}

\affil{$^1$Astrophysikalisches Institut Potsdam, An der Sternwarte 16,
  14482 Potsdam, Germany; \\
{\tt rklessen@aip.de}}

\affil{$^2$Centro de Radioastronom\'ia y Astrof\'isica,
UNAM. Apdo. Postal 72-3 (Xangari), Morelia, Michoc\'an 58089, M\'exico \\ 
{\tt j.ballesteros@astrosmo.unam.mx}; {\tt e.vazquez@astrosmo.unam.mx;
c.duran@astrosmo.unam.mx}}

\slugcomment{Draft date: \today}

\lefthead{Ballesteros-Paredes, Klessen \& V\'azquez-Semadeni}

\righthead{Hydrostatic disguise. II.}

\begin{abstract}
  
  We investigate the velocity structure of protostellar cores that
  result from non-magnetic numerical models of the gravoturbulent
  fragmentation of molecular cloud material.  A large fraction of the
  cores analyzed are ``quiescent''; i.e., have non-thermal linewiths
  smaller or equal to the thermal linewidth. Specifically, about 23\%
  of the cores have subsonic turbulent line-of-sight velocity
  dispersions $\sigma_{\rm turb} \le c_{\rm s}$.  A total of 46\% are
  ``transonic'', with $c_{\rm s} < \sigma_{\rm turb} \le 2 c_{\rm s}$.
  More than half of our sample cores are identified as ``coherent'',
  i.e., with $\sigma_{\rm turb}$ roughly independent of column
  density. Of these, about 40$\,$\%\ are quiescent, 40$\,$\%\ are
  transonic, and 20$\,$\%\ are supersonic.

  The fact that dynamically evolving cores in highly supersonic
  turbulent flows can be quiescent may be understood because cores lie
  at the stagnation points of convergent turbulent flows, where
  compression is at a maximum, and relative velocity differences are
  at a minimum. The apparent coherence may be due, at least in part,
  to an observational effect related to the length and concentration
  of the material contributing to the line.
  
  In our simulated cores, $\sigma_{\rm turb}$ often has its local
  maximum at small but finite offsets from the column density maximum,
  suggesting that the core is the dense region behind a shock. Such a
  configuration is often found in observations of nearby molecular
  cloud cores, and argues in favor of the gravoturbulent scenario of
  stellar birth as it is not expected in star-formation models based
  on magnetic mediation.
  
  A comparison between the virial estimate $M_{\rm vir}$ for the mass
  of a core based on $\sigma_{\rm turb}$ and its actual value $M$
  shows that cores with collapsed objects tend to be near
  equipartition between their gravitational and kinetic energies,
  while cores without collapsed objects tend to be gravitationally
  unbound, suggesting that gravitational collapse occurs immediately
  after gravity becomes dominant.
  
  Finally, cores in simulations driven at large scales are more
  frequently quiescent and coherent, and have more realistic ratios of
  $M_{\rm vir}/M$, supporting the notion that molecular cloud
  turbulence is driven at large scales.

\end{abstract}


\section{Introduction}\label{intro}

Understanding the processes that lead to the formation of stars is
one of the fundamental challenges in theoretical astrophysics. It is
well known that stars form in dense cores within molecular
clouds, but the physical processes that control the formation of
low-mass stars within these cores are not well understood yet.

The traditional scenario assumes that low-mass protostellar cores are
in quasi-static equilibrium supported against gravitational collapse
by a combination of magnetic and thermal pressures (see, e.g., Shu,
Adams \& Lizano 1987). A core forms stars once
magnetic support is lost through a process called ambipolar diffusion.
Neutral gas particles slowly drift through the ions which are held up
by the magnetic field, allowing the core to eventually attain a critical
mass-to-flux ratio.  Then the gravitational energy exceeds the
magnetic energy and collapse sets in from the inside-out.

The theory of gravoturbulent star formation (see, e.g., the reviews by
V\'azquez-Semadeni et al.\ 2000; Larson 2003; Mac Low \& Klessen 2004,
and references therein), on the other hand, suggests that clouds and
cores are formed by compressible motions in the turbulent velocity
field of their environment (e.g.\ von Weizs{\"a}cker 1951; Sasao 1973;
Hunter \& Fleck 1982; Elmegreen 1993; Padoan 1995;
Ballesteros-Paredes, Hartmann \& V\'azquez-Semadeni 1999a;
Ballesteros-Paredes, V\'azquez-Semadeni \& Scalo 1999b; Klessen et
al.\ 2000; Padoan et al.\ 2001a; Heitsch et al.\ 2001;
V\'azquez-Semadeni et al.\ 2004).  Those cores with an excess of
gravitational energy collapse rapidly to form stars, while the others
with sufficiently large internal or kinetic energies re-expand once
the turbulent compression subsides.

Observational evidence suggests that low-mass stars form from
molecular cloud cores with column density profiles that often resemble
those of Bonnor-Ebert\footnote{Ebert (1955) and Bonnor (1956) describe
  the equilibrium density structure of isothermal gas spheres confined
  by an external pressure as solution of the Lane-Emden equation.}
equilibrium spheres (Alves, Lada, \& Lada 2001, see also the review by
Andr{\'e}, Ward-Thompson, \& Barsony 2000), and with velocity
dispersions that are small, i.e.\ transonic, or even subsonic (Myers
1983; Barranco \& Goodman 1998; Goodman et al.\ 1998; Jijina, Myers,
\& Adams 1999; Caselli et al.\ 2002; Tafalla et al.\ 2004).  For this
reason such cores are often termed ``quiescent''.  Moreover, if the
measured line-of-sight (l.o.s) velocity dispersion of a core is
independent of column density towards the maximum, then it is called
``coherent'' (Barranco \& Goodman 1998).  In the scenario of
magnetically mediated star formation (Shu et al.\ 1987) these
structures are explained as consequences of the quasistatic
contraction process.  In the gravoturbulent theory, however,
protostellar cores are transient features naturally generated by the
dynamical flow in the cloud. In order to test this theory, it is
necessary to show that these fluctuations exhibit properties similar
to those of the observed cores.

Several groups have now began to study core properties in numerical
simulations of gravoturbulent cloud fragmentation (e.g., Ostriker,
Stone \& Gammie 2001; Padoan et al.\ 2001a,b; Ballesteros-Paredes \&
Mac Low 2002; Gammie et al.\ 2003; Li et al.\ 2004; Schmeja \& Klessen
2004; Jappsen \& Klessen 2004; Tilley \& Pudritz 2004;
V\'azquez-Semadeni et al.\ 2004).  In particular, Ballesteros-Paredes,
Klessen \& V\'azquez-Semadeni (2003, hereafter called Paper~I)
demonstrated that indeed transient, dynamic cores have an
angle-averaged column density structure that often resembles that of
hydrostatic Bonnor-Ebert profiles. This analysis was based on
numerical calculations by Klessen, Burkert \& Bate (1998), Klessen \&
Burkert (2000, 2001), and Klessen et al.\ (2000), and applied a
fitting procedure similar to that used by Alves, Lada, \& Lada (2001).

In this paper we focus on the velocity structure of protostellar
cores, and compare with the data available for observed quiescent,
low-mass cores.  In \S\ref{simulations:sec} we summarize the main
features of the numerical models used, and explain how we analyze the
density and velocity structure.  In \S\ref{coherent:sec} we show that
the density fluctuations that we identify with protostellar cores
often have very small and nearly spatially constant turbulent l.o.s
velocity dispersion, even though they are produced by highly turbulent
supersonic flows. In \S\ref{energies:sec} we discuss the energy budget
of the cores.  Finally, in \S\ref{discussion:sec} we summarize and
interpret our results in terms of the gravoturbulent fragmentation
model of star formation.

\section{Numerical simulations and core sample}\label{simulations:sec}

An important prerequisite for adequately describing the density and velocity
structure of cores in numerical models of gravoturbulent molecular cloud
evolution is the ability to resolve high density contrasts at arbitrary
locations within the cloud.  Smoothed particle hydrodynamics (SPH; see Benz
1990, Monaghan 1992) is probably the best method currently available for this
purpose.

The properties of our numerical scheme and resolution issues in the
context of gravoturbulent fragmentation have been extensively
discussed in Paper~I (see also Klessen et al.\ 2000, or Klessen 2001).
Once the density contrast in the center of a collapsing cloud core
exceeds a density contrast of about $10^4$ a ``sink'' particle is
created (Bate et al.\ 1995). It replaces the central high-density
region and has the ability to accrete further infalling material while
keeping track of mass and linear and angular momentum.  However,
the internal structure of the sink particle is not resolved. With a
diameter of about 600 AU it fully encloses the star/disk system 
expected to form roughly 1000 years after the critical density for
sink particle formation is reached (Wuchterl \& Klessen 2000).

The numerical resolution limit of our numerical scheme is determined
by the Bate \& Burkert (1996) criterion, which is sufficient for the
highly nonlinear fluctuation spectrum considered here. This is
confirmed by resolutions studies with up to $10^7$ SPH particles (see
Jappsen et al.\ 2004). It should be noted, however, that the Bate \&
Burkert (1996) criterion may not be sufficient for adequately
following the growth of linear perturbations out of quasi-equilibrium,
as was suggested for the case of rotationally supported disks by
Fisher, Klein, \& McKee (private communication).

We analyze two models, one labeled \lsd, in which turbulence is driven
on large scales, of wavelength $\lambda \approx 1/2$ of the
computational box, and the other labeled \ssd, in which energy is
injected on smaller scales, of $\lambda\approx 1/8$ of the box. We
consider the system at an evolutionary stage when 5\% of the available
gas mass is accumulated in collapsed cores. Note that in Paper~I we also
studied cores from a contracting Gaussian density field (GC) without
turbulence. Since in this paper we focus on the turbulent velocity
structure, that simulation is not considered here.

To identify cloud cores we use the 3-dimensional clump-finding
algorithm introduced in Appendix A of Klessen \& Burkert (2000).  We
then project a cubic subregion of the full computational volume
centered around the core along the three principal axes, and compute
the column density $N$ and the total, turbulent-plus-thermal l.o.s
velocity dispersion $\sigma_{\rm los}$ of each core. We take
$\sigma_{\rm los}^2 = \sigma_{\rm turb}^2 + c_{\rm s}^2$, where the
turbulent velocity dispersion $\sigma_{\rm turb}$ is obtained as the
mass-weighted standard deviation of the velocity field in each
line-of-sight along the projection axis, and $c_{\rm s}^2$ is the
sound speed of the mean particle ($\mu = 2.3\, m_{\rm H}$).  This assumes
optically thin emission throughout the subcube, and gives us a direct
estimate of the true physical state of the system. An in-detail
comparison with observations requires to consider specific molecular
emission lines tracing various density regimes and to take optical
thickness effects into account.

Since each projection in general gives different values of $N$, core
size $R$ and $\sigma_{\rm los}$, we treat each projection as an
independent case.  We have increased the number of analyzed cores in
comparison to Paper~I by taking the first 200 cores identified in each
of the LSD and SSD simulations. For the quiescence and coherence
studies we only consider ``starless'' cores and exclude those with
collapsed central regions, i.e.\ with a sink particle in their
interior.  For the energy budget analysis, however, we do include
cores with sinks, as they allow for comparison with observations of
cores containing young stellar objects.  In order to avoid repetition,
we analyze fields with multiple cores detected by the clumpfinding
algorithm only once.  This procedure yields a sample of 44 cores for
the LSD model, and 101 for SSD without central collapsed object, plus
15 and 10 fields with sink particles for LSD and SSD models,
respectively.  Altogether, we analyze $(44 + 101) \times 3 = 435$
column density maps without, and $(15+10) \times 3 = 75$ column
density maps with sink particles.  Adopting the same physical scaling as in
Paper~I, the maps we consider cover $0.154\,$pc by $0.154\,$pc. The
mean density in the simulation is $n({\rm H_2}) = 3.3 \times
10^3\,$cm$^{-3}$, the total mass in the simulation corresponds to
$\sim 700\,M_{\odot}$, and the speed of sound is $c_{\rm s} =
0.2\,$km$\,$s$^{-1}$.

\section{Results}\label{results:sec}

\subsection{Quiescent and coherent cores from gravoturbulent fragmentation} \label{coherent:sec} 

In the following, we consider a core being {\em quiescent} if its
projected non-thermal velocity dispersion $\sigma_{\rm turb}$ is
smaller or equal to the thermal sound speed ($\sigma_{\rm turb} \le
c_{\rm s}$), within the half central surface density contour, i.e.\ if
the non-thermal component to the observed linewidth does not exceed
the thermal line broadening. We call a core {\em coherent} if its
velocity dispersion is independent of column density again for $N$
larger than half the peak value $N_{\rm max}$, or in other words, if
$\sigma_{\rm turb}$ is roughly independent of the offset from the core
center.

\begin{figure} 
\begin{center}
\includegraphics[width=9cm]{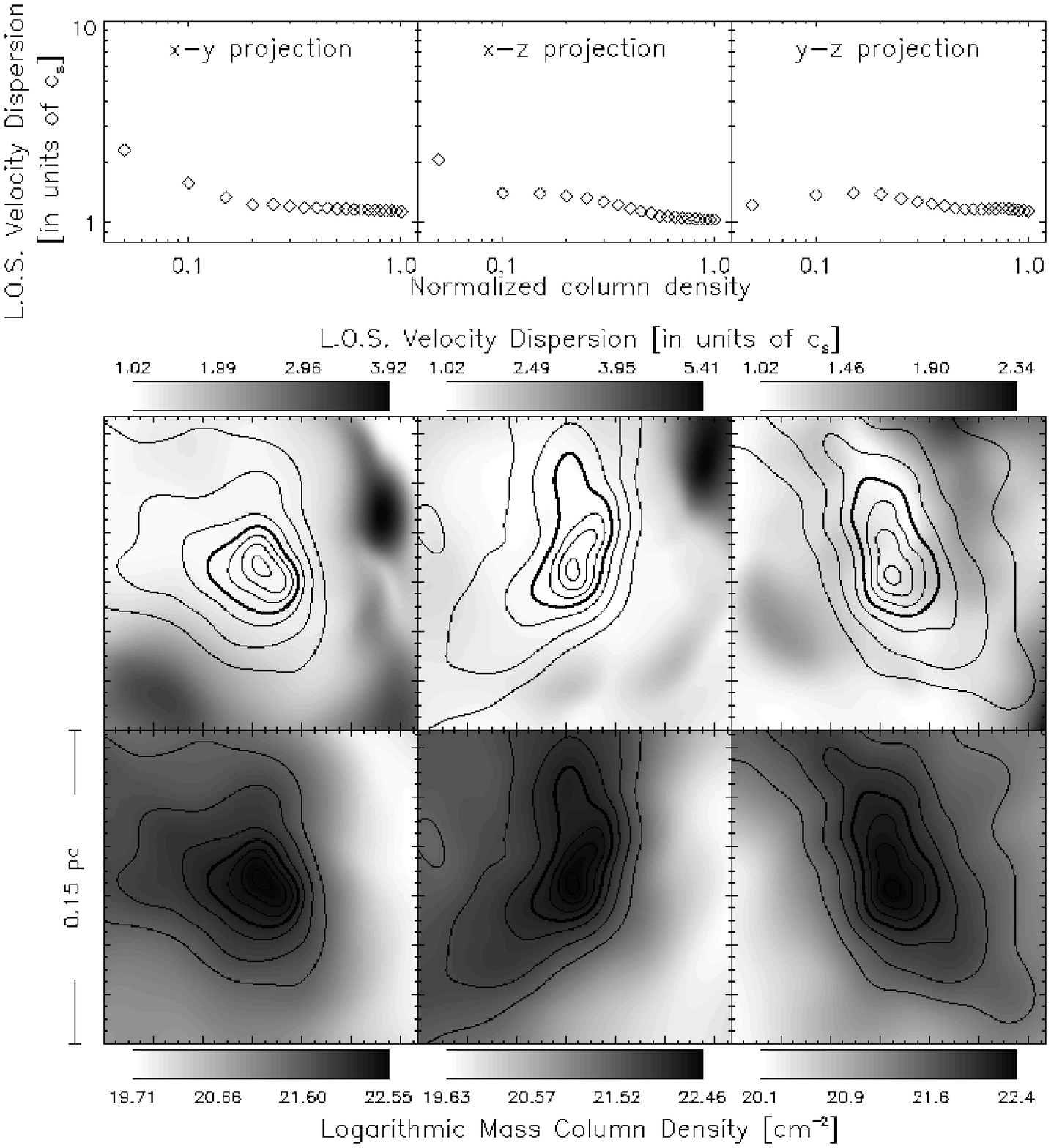}

\vspace{0.3cm}

\includegraphics[width=9cm]{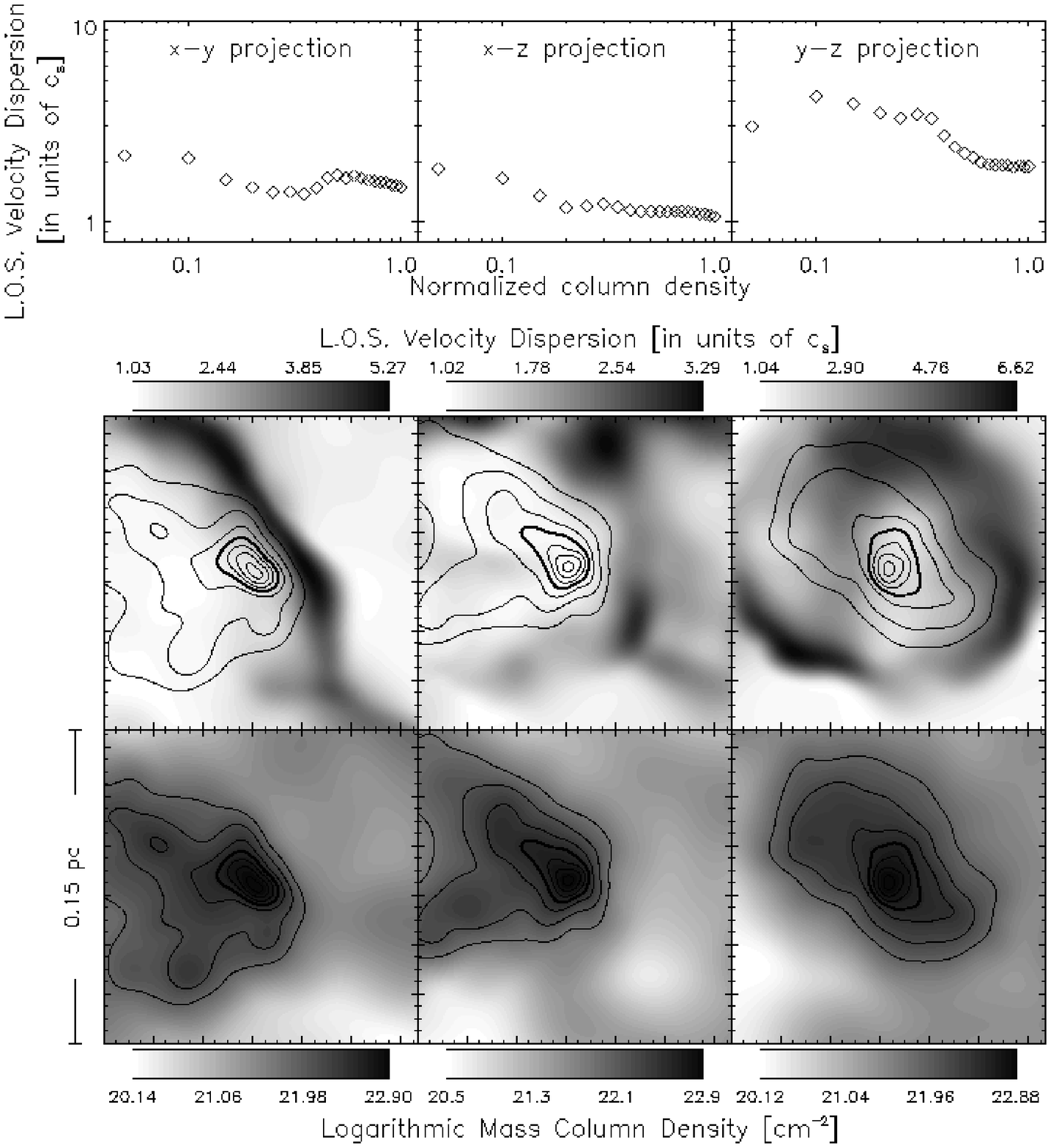}
\end{center}

\vspace{-0.7cm}
\caption{Two selected cores from model LSD in their three projections.
  For each core, the {\em lower panels} give grayscale maps of
  logarithmic column density $N$. The {\em middle panels} show the
  total, i.e.\ turbulent plus thermal, line-of-sight velocity
  dispersion $\sigma_{\rm los} = \sqrt{\sigma^2_{\rm turb} + c_s^2}$
  superimposed on the column density contours. In the {\em upper
    panels} we plot $\sigma_{\rm los}$   against normalized column
  density $N$. We normalize $\sigma_{\rm los}$ to the thermal sound
  speed $c_{\rm s}$ as indicated by the upper grayscale bar), and we
  plot $N$ in logarithmic units as indicated by the lower grayscale
  bar.  For better orientation, we also indicate the density structure
  with contour lines in the lower and middle panel. Contour levels are
  drawn in linear scaling at 10$\,$\%, 20$\,$\%, 35$\,$\%, 50$\,$\%,
  65$\,$\%, 80$\,$\%, 95$\,$\% of the peak value $N_{\rm max}$. The
  50$\,$\%-isocontour is marked with a thicker line.
\label{vel_disp_1:fig}}
\end{figure}

\begin{figure} 
\begin{center}
\includegraphics[width=10cm]{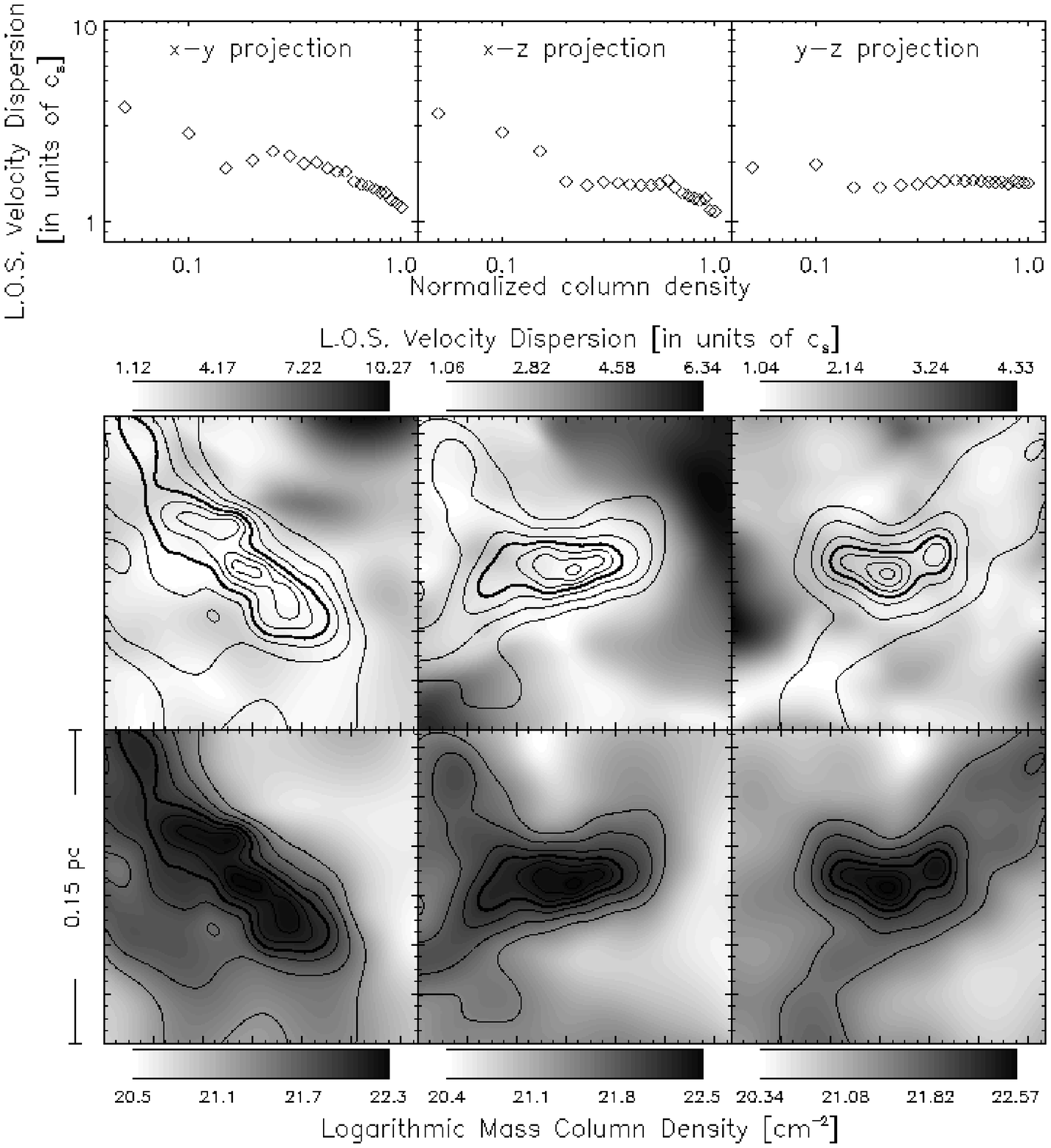}

\vspace{0.3cm}

\includegraphics[width=10cm]{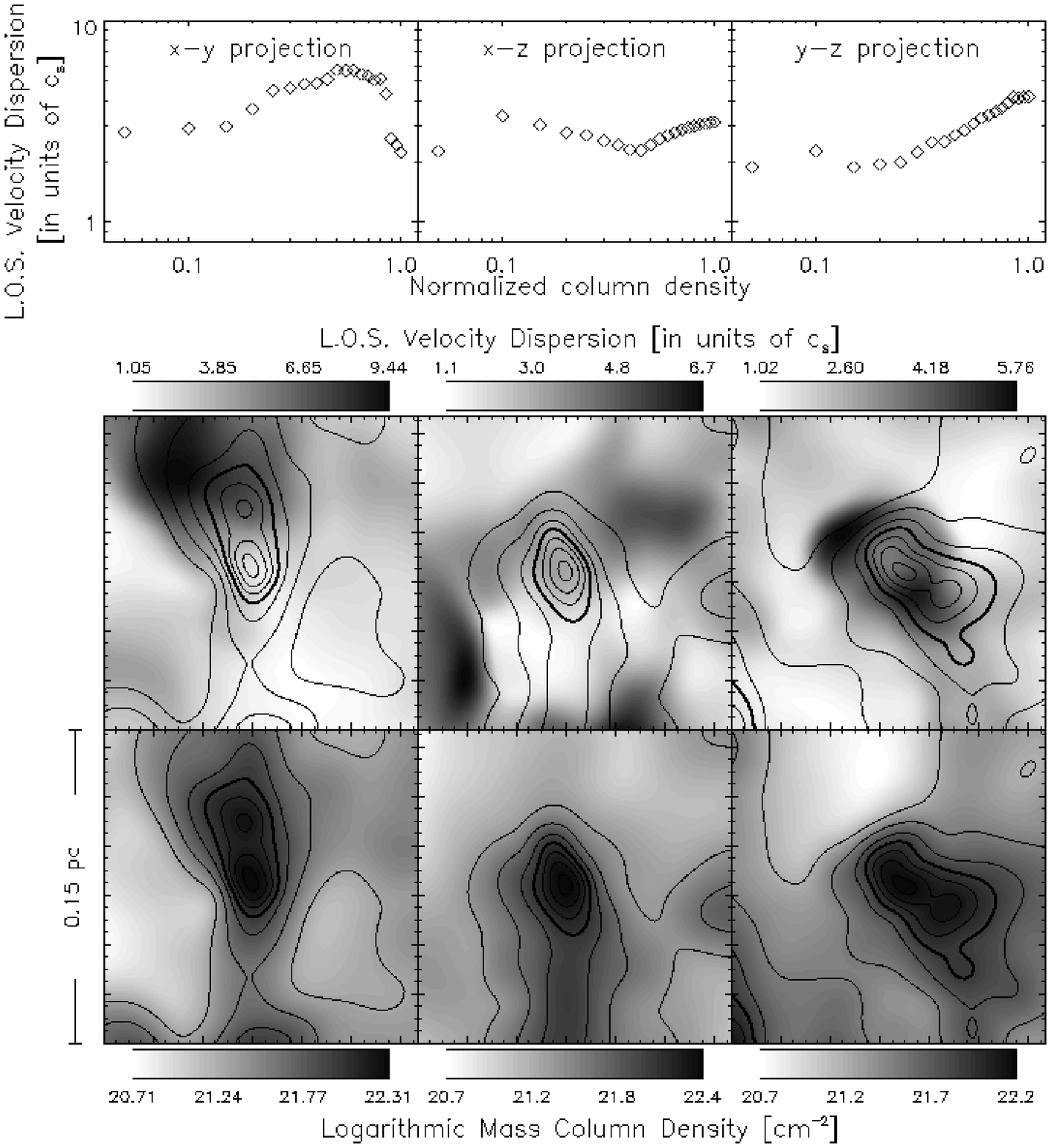}
\end{center}

\caption{Two selected cores  from
  SSD  in their three projections. Notation and scaling is identical
  to Figure \ref{vel_disp_1:fig}.
\label{vel_disp_2:fig}}
\end{figure}

In Figures \ref{vel_disp_1:fig} and \ref{vel_disp_2:fig} we illustrate
the properties of four different molecular cloud cores in our models.
We show maps of column density $N$, and maps of the l.o.s velocity
dispersion $\sigma_{\rm los}$ for each projection.  To check for
coherence, we also plot $\sigma_{\rm los}$ versus the normalized
column density $N/N_{\rm max}$.  In this particular case, we average
$\sigma_{\rm los}$ in bins of 0.1 with respect to the normalized
column density.  This allows for a direct comparison, e.g., with
Fig.~4 in Barranco \& Goodman (1998), who plot the velocity dispersion
against the antenna temperature using similar binning.  The cores are
chosen to span a wide range of morphological and structural
appearance.  Figure \ref{vel_disp_1:fig} shows two cores from model
LSD. The top one is a smooth and roundish object that is coherent as
well as quiescent in each of its projections. Its density as well as
kinematic structure closely resembles that of the two ``thermal''
cores L1498 and L1517B in Taurus, studied in great detail by Tafalla
et al.\ (2004). In the inner parts the derived velocity dispersion is
almost entirely determined by thermal motion only. The lower core in
Figure~\ref{vel_disp_1:fig} has $xy$- and $xz$-projections that are
classified as subsonic-coherent, but it appears transonic-coherent in
the $yz$-map.  Recall that, similarly to the standard observational
procedures, we consider only column densities above half of the peak
value.  In contrast to the upper core, it is clearly cometary-shaped as
result of highly anisotropic ram pressure.  The external flow coming
from the right-hand side (in the $xy$- and $xz$-projection) is
abruptly stopped at the surface of the core, leading to a noticeable
increase of the velocity dispersion there. In
Fig.~\ref{vel_disp_2:fig} we present cores from model SSD. The first
case is coherent only in the $yz$-map, in the other two projections
$\sigma_{\rm los}$ varies too strongly with location and consequently
with $N$.  The last one represents the subset of cores that are
neither quiescent nor coherent in any of their projections.

The fact that high-density clumps in turbulent molecular clouds are created by
convergent flows and thus are transiently bounded by ram pressure has
observational consequences.  As illustrated in
Figures~\ref{vel_disp_1:fig} and ~\ref{vel_disp_2:fig}, 
in the simulations we often find localized maxima of $\sigma_{\rm
los}$ in the low column-density gas at the outskirts of the core,
suggestive that the core is actually a dense post-shock region, with the
localized maximum of $\sigma_{\rm los}$ signaling the shock position.
Such a configuration
is also often seen in observations of actual molecular cloud cores (e.g., 
Barranco \& Goodman 1998; Goodman et al.\ 1998; Caselli et al.\ 2002;
Tafalla et al.\ 2004).
This important observational feature is thus naturally explained by
the theory of gravoturbulent star formation.  It should be noted, that
the ``standard'' scenario of magnetically mediated star formation
treats the turbulent nature of the cores' surroundings in a rather
{\it ad-hoc} fashion, through the consideration of micro-turbulent
motions providing an isotropic pressure that increases with decreasing
density (e.g., Lizano \& Shu 1989, Myers \& Fuller 1992). In this
case, the increase of $\sigma_{\rm los}$ outside  the core should
in general be roughly isotropic, contrary to observed localized maxima
of the velocity dispersion (see, e.g., maps in Barranco and Goodman
1998; Goodman et al.\ 1998, Caselli et al.\ 2002).

The origin of the coherence, on the other hand, is not so clear. We
speculate that it may arise from the fact that the density-weighted
path length of gas contributing to the emission has a minimum at the
core center, as illustrated in Figure 4 of Goodman et al.\ (1998).
This contributing length is thus stationary with respect to small
offsets in the plane of the sky. If the linewidth in a pencil beam is
due essentially to density-weighted velocity differences sampled along
the line-of-sight, then it should exhibit the same behavior. Because
our analysis is based on information from all the available gas, the
quiescent and coherent nature of some cores in our sample is
inequivocal evidence of the abscence of large velocities in their
interior.  This suggests that either these cores have not developed
gravitational collapse motions, or else are at the earliest stages of
collapse, with velocities still being small.  Whether this is the
correct interpretation of observed line maps requires detailed
radiative transfer calculations for various tracer molecules, and we
plan to test this speculation in detail in a forthcoming paper.

 From Figures \ref{vel_disp_1:fig} and \ref{vel_disp_2:fig} and the
statistics of the complete core sample\footnote{Note that choosing
a different time for analysis in our simulations (cf. \S 2), or
different parameters for the simulations, would probably alter to some
extent the statistics presented in this paper.  Thus, the precise
fractions of coherent and quiescent cores should not be taken
literally.  The fundamental result is that a substantial fraction of
the cores in the simulations analyzed here, at an arbitrary time in
their evolution, are quiesent and/or coherent.} (Table  
\ref{tab:coherence}), we note several issues. First, as already
discussed in Paper~I, the inferred properties of cores may vary
considerably between different projections.  For example, $\sigma_{\rm
  turb}$ may vary by as much as a factor of 2 to 3.  Second, in our
turbulent models, roughly 60\% of all cores can be identified as being
coherent by visual inspection. About 40\% of these are quiescent or
subsonic, 40\% are transonic, and about 20\% are supersonic.  Third,
we note a clear distinction between the LSD and SSD model, the former
having a larger fraction of coherent cores, most of which are
furthermore quiescent. Instead, the SSD model produces a larger number
of cores that do not qualify as being coherent, and those that do are
mostly transonic or supersonic. Thus, the LSD model compares better to
observations than the SSD model because observed cores often appear
quiescent and coherent. However, persuasive observational statistics
is still lacking. This argues in favor of clouds and their cores being
driven from large scales (see also the discussion in Ossenkopf \&
Mac~Low 2002).

\begin{figure} 
\begin{center}
\includegraphics[width=10cm]{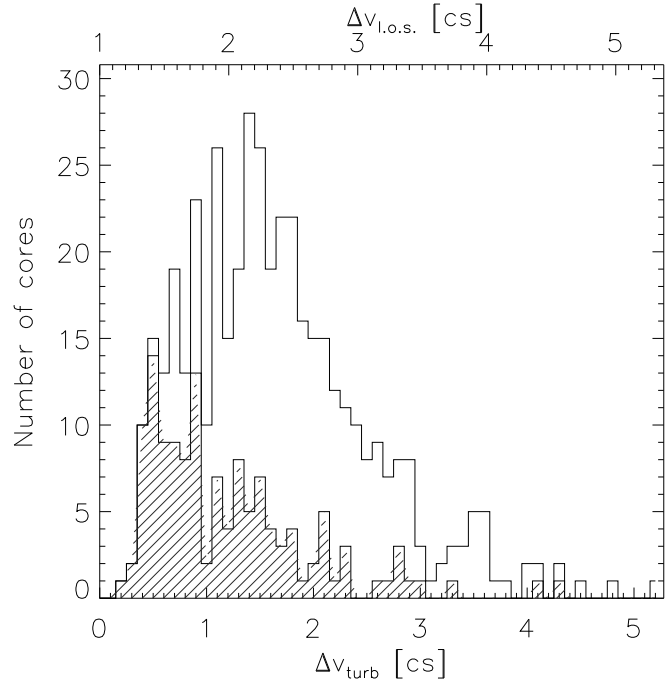}
\end{center}
\caption{Histogram of the mean line-of-sight velocity dispersion
  $\sigma_{\rm los}$ inside half of the maximum column density.
  Lower abscissa gives the scaling for the turbulent l.o.s velocity
  dispersion $\sigma_{\rm turb}$, and upper abscissa denotes total
  l.o.s velocity dispersion $\sigma_{\rm los}$, given by
  $\sqrt{\sigma^2_{\rm turb} + c_s^2}$. The distribution in model
  LSD is given by the hatched thick-line histogram, the histogram of
  model SSD is drawn with thin line.
\label{histo:fig}}
\end{figure}

To estimate the fraction of cores with subsonic velocities,
Fig.~\ref{histo:fig} shows a histogram of the mean velocity dispersion
inside the lowest contour (above half of the peak column density). For
the whole core sample (models LSD and SSD combined) about 23$\,$\% of
all cores are quiescent in the strict sense (i.e.\ with $\sigma_{\rm
  turb} \le c_{\rm s}$), while 46$\,$\% are still transonic with
$c_{\rm s} < \sigma_{\rm turb} \le 2 c_{\rm s}$.  For the preferred
model LSD, roughly 50\% of them are coherent and subsonic, 15\% are
coherent and transonic, only 6\% are coherent and supersonic, and the
remaining 29\% are not found to be coherent.

We stress that all of this occurs in turbulent flows with rms Mach
numbers as high as 6.  In fact, this is a natural consequence of the
turbulent energy cascade. The velocity field becomes progressively
more auto-correlated towards small scales leading to the well-known
line width--size relation in Galactic molecular clouds (Larson 1981).
In interstellar turbulence, there is thus always a lengthscale at
which the flow turns from supersonic to subsonic (Padoan 1995;
V\'azquez-Semadeni, Ballesteros-Paredes, \& Klessen 2003a).  This does
not necessarily imply that the dissipative regime of the turbulence
has been reached, nor that there is a unique inner scale, but only
that, on average, at this scale in a particular flow, the cascade
turns into an incompressible one.  In Galactic molecular clouds, this
happens at roughly 0.05--0.1$\,$pc (e.g., Larson 1981; Myers 1983),
which is the typical size of cold cloud cores.  Again, this does not
imply these cores are quasistatic or long-lived. On the contrary,
cores in our simulations are always out of equilibrium.  They are
created by a transient turbulent compression that eventually subsides,
at which point the core is left in an unbalanced state. If the core
crosses the ``border'' of gravitational instability, then it
immediately proceeds to collapse.  Otherwise, it re-expands and merges
back into its surroundings in times slightly longer than the local
free-fall time (V\'azquez-Semadeni et al.\ 2004). Moreover, cores can
be disturbed, destroyed or re-compressed by interaction with
neighboring turbulent fluctuations (e.g., Klessen et al.\ 2000;
Klessen 2001).

\subsection{Energy balance in protostellar cores}\label{energies:sec}

Observed protostellar cores are often said to be in ``virial
equilibrium'' (e.g. Myers 1983; Myers \& Goodman 1988a,b), because the
measured and the inferred ``virial'' values of certain physical
variables are comparable. Recent studies have focused on the
comparison between of the observationally estimated mass $M$ and the
virial mass $M_{\rm vir}$, defined as
\begin{equation}
M_{\rm vir} (M_\odot) \equiv 210 \ R({\rm pc}) \  \Delta
v_{\rm eff}^2 ({\rm km^2\,s^{-2}})
\label{eq:virial_mass}
\end{equation}
(see, e.g., Caselli et al.\ 2002; Tachihara et al.\ 2002) for a
uniform density sphere of radius $R$ and ``effective'' linewidth
$\Delta v_{\rm eff}$.  Because most observations are based on the
emission from tracer molecules heavier than the mean particle mass in
the gas ($\mu = 2.3 \,m_{\rm H}$), the thermal contribution to
 $\Delta v_{\rm eff}$ must be
calculated by substracting the one from the tracer molecule and adding
a fictitious  contribution 
from the mean molecule (see, e.g., eq.\ 7 of Caselli et al.\ 2002).
Given the definition of the line-of-sight velocity dispersion $ \sigma_{\rm los}$ in \S\ref{simulations:sec}, for our theoretical data this
``effective'' linewidth is computed as $\Delta v_{\rm eff}=(8
\ln{2})^{1/2}\sigma_{\rm los}$.

\begin{figure} 
\begin{center}
\includegraphics[width=14cm]{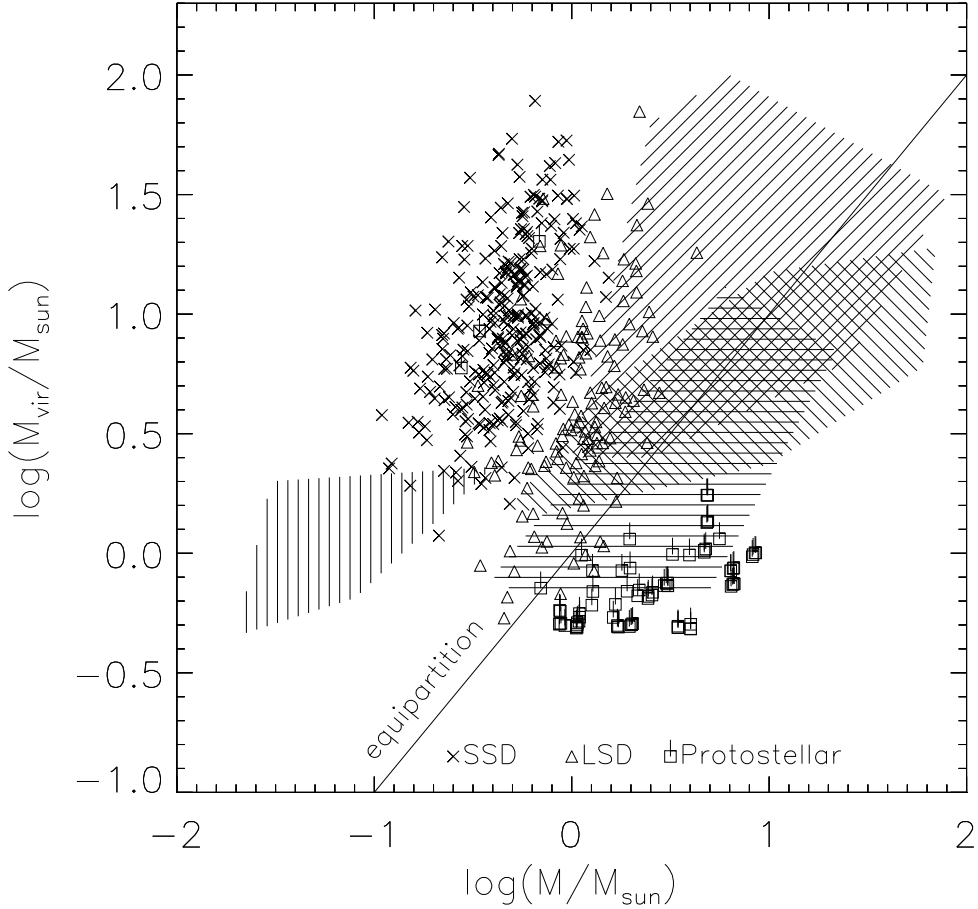}
\end{center}
\caption{
  Estimated virial mass $M_{\rm vir}$ plotted against actual mass $M$
  for all analyzed cores in the simulations. {\em Crosses} denote
  cores in the small scale turbulence model (SSD) and {\em triangles}
  denote cores in the large scale turbulence model (LSD). {\em Tailed
    Squares} indicate the lower limits on the estimates on $M_{\rm
    vir}$ for cores in LSD with protostellar objects (sink particles)
  in their interior. Note that all three projections for each core are
  plotted independently.  The identity, $M_{\rm vir} = M$, is given by
  the solid line. We also indicate the parameter space covered by
  cores in the observational surveys by Morata et al.\ (2005) with
  vertical lines, by Onishi et al.\ (2002) with horizontal lines, by
  Caselli et al.\ (2002) with $-45^{\circ}$ lines, and we plot the
  starless cores in Tachihara et al.\ (2002) with $+45^{\circ}$ lines.
\label{energies:fig}}
\end{figure}

In Fig.~\ref{energies:fig} we plot $M_{\rm vir}$\ against $M$. The
actual core mass $M$ is calculated by integrating the column density
within the half-maximum isocontour of the column density maps. The
diagonal line denotes equipartition, $M_{\rm vir} = M$.  Crosses
(triangles) denote cores in the SSD (LSD) model. Cores with protostars
(i.e., in our numerical sheme with a sink particle in their center)
are indicated by tailed squares. Note that their virial mass estimates
$M_{\rm vir}$ are lower limits.  Although the sink particle carries
the correct mass, there is no information on its internal velocity
structure, because it is not resolved in the SPH code.  Thus, the
velocity dispersion calculated in fields with sink particles is an
underestimate.  However, this underestimate does not appear to be too
serious, since at the evolutionary stage we consider, the mass in the
central protostar is just a small fraction of the overall core mass.
Moreover, a similar underestimate is likely to occur in the
observations due to depletion and optical thickness effects in the
dense, collapsing gas in the central regions of real cores. To allow
for direct comparison with observational data, we also indicate in
Figure ~\ref{energies:fig} the regions covered by the cores in the
surveys of Morata et al.\ (2004, vertical lines), Onishi et al.\ 
(2002, horizontal lines), Caselli et al.\ (2002, lines tilted by
$-45^{\circ}$), and starless cores in Tachihara et al.\ (2002, lines
tilted by $+45^{\circ}$).

With the above considerations in mind, several points are worth
noting.  First, cores from the large-scale turbulence model LSD
populate a different region in $M_{\rm vir}$--$M$ parameter space than
their counterparts in the small-scale turbulence model SSD.  The
former tend to have somewhat lower line-of-sight velocity dispersion
$\sigma_{\rm los}$ (implying lower virial masses) and at the same time
larger actual masses. As a result, they are closer to equipartition
than cores in SSD.  The velocity field in the LSD model is dominated
by large-scale shocks that are very efficient in sweeping up molecular
cloud material, thus creating massive coherent density structures that
more frequently exceed the critical mass for gravitational collapse.
Therefore cloud cores in the LSD model predominantly form in a
clustered and coeval mode from gas that is largely Jeans unstable (see
Klessen 2001). On average they have higher density contrast than their
SSD counterparts, and their energy budget is more influenced by
self-gravity.  The LSD cores fall almost completely in the low-mass
regions of the Onishi et al.\ (2002), Caselli et al.\ (2002) and
Tachihara et al.\ (2002) samples, suggesting that the model reproduces
well the physics of the lower-mass cores in these regions, while the
production of higher- or lower-mass cores by the simulations probably
requires a different normalization and/or varying the mass content in
the simulation (by varying the number of Jeans masses in the box).

On the other hand, shock-generated clumps in the SSD model tend to
form at random locations and at random times. On average they have a
smaller density contrast and smaller masses.  By the same token, their
velocity dispersions are larger, giving thus larger virial masses
$M_{\rm vir}$.  These cores tend to be more dominated by their
internal velocity dispersion rather than by self-gravity, as reflected
by the fact that they exibit larger departures from equipartition
(typically by factors $10$ to $100$), which appear inconsistent with
observational estimates of this balance, except for the strongly
sub-virial cores presented discussed Morata et al.\ (2004). 

Second, the starless cores in our sample exhibit an excess of their
virial mass.  This result is consistent with recent observations by
Tachihara et al.\ (2002) and Morata et al.\ (2004), who show that for
cores without central protostar the virial mass estimates usually
exceed the actual mass.  Although in the Tachihara et al.\ (2002)
sample departures from equipartition smaller (less than a factor of
10), Morata et al.\ (2004) show cores that depart from equipartition
by factors of 30.  On the other hand, the cores classified as starless
by Caselli et al.\ (2002) tend to be in equipartition.  The above
results suggest that some of these may actually contain collapsed
objects.  Observational evidence of this possibility has been recently
given by new observations with the Sptizer Space Telescope.  On one
hand, Reach et al.\ (2004) have detected 8 embedded sources (Class
0/I) in a small field centered on a single globule in Tr 37 where only
one IRAS source was detected.  Similarily, Young et al.\ (2004), show
that L1014, a dense core previously thought to be starless actually
shows evidence of containing an embedded source.  Thus, if this
phenomenon is common, then many apparently starless cores that are
near equipartition may actually contain collapsed objects.
Furthermore, as pointed out by Young et al.\ (2004), this would also
suggest that traditional estimates of pre-stellar core lifetimes may
be overestimated.  This again supports the idea that not all density
peaks in self-gravitating turbulent fields necessarily collapse, but
that collapse occurs rapidly once equipartition is reached
(V\'azquez-Semadeni et al.\ 2004).

A third point to make is that cores with central protostars (sink
particles) in our models tend to be more massive than cores without,
in agreement with the observational situation that cores with stars
tend to be more massive than starless cores (Caselli et al.\ 2002;
Tachihara et al.\ 2002)

Finally, we note, that even though some cores in the model lie close
to the identity line $M_{\rm vir} = M$, it does not imply virial {\em
  equilibrium}. This requires the second derivative of the moment of
inertia to vanish (see, e.g., Ballesteros-Paredes 2004 and references
therein).  Our cores are not static, but instead are constantly
evolving, and thus they are in general {\em out} of equilibrium.
Reaching hydrostatic equilibrium in a turbulent molecular cloud
environment is extremely difficult, and requires strongly idealized
conditions that are not met in the interstellar gas.  The condition
$M_{\rm vir} = M$, or equivalently $\sigma_{\rm los} \approx
\sigma_{\rm vir}$, simply reflects equipartition between the
volume-averaged kinetic energy and self-gravity, as occurs precisely
at the verge of gravitational collapse.

Altogether, our calculations support the following evolutionary
sequence. Initially, cloud cores are generated by transient
compressive turbulent motions. In this phase their energy budget is
dominated by the external ram pressure. The compression causes their
internal and gravitational energies to increase. If they accumulate
enough mass, or reach sufficient density contrast, they may become
gravitationally unstable and quickly go into collapse. The transition
to the stage when self-gravity dominates the evolution is
characterized by approximate energy equipartition. If self-gravity
never becomes that important, the cores are left with an excess of
internal energy after the external compression subsides, and may
re-expand within a few free-fall times.

\section{Summary and Conclusions} \label{discussion:sec}

In the emerging picture of gravoturbulent star formation (see, e.g.,
the review by Mac Low \& Klessen 2004, and references therein), the
structure of Galactic molecular clouds is determined by compressible
supersonic turbulence.  High-density cores build-up at the stagnation
points of locally convergent flows.  Some of those cores may become
gravitationally unstable and go into collapse to form stars, while
others will simply redisperse into the ambient medium (Sasao 1973;
Hunter \& Fleck 1982; Elmegreen 1993; Ballesteros-Paredes et al.\ 
1999a; Padoan et al.\ 2001a; Klessen et al.\ 2001; Padoan \& Nordlund
2002; V\'azquez-Semadeni et al.\ 2004).

Our analysis demonstrates that a considerable fraction of the cores in
supersonic turbulent flows can be identified as being ``quiescent''
(i.e., sub- or transonic) and ``coherent'' (i.e., with roughly
constant velocity dispersion across the central parts of the core)
despite the fact that they are embedded and formed in a highly
dynamical environment. These cores are quiescent, because they form at
the stagnation points of the flow where the compression is at a
maximum and the relative velocity differences are at a minimum.  The
origin of coherence is not so clear, but we speculate that it may be
caused by projection, because the line-of-sight length of the matter
contributing to the line profile has a minimum at the core center,
and small offsets from the center therefore cause little variation in
the observed line width.

Molecular cloud cores that harbor protostars in their interior (as
identified by the presence of sink particles in our models) are
characterized by having $M_{\rm vir}<M$, but for those objects our
estimates to $M_{\rm vir}$ are lower limits to the true value.  On the
other hand, most of the cores without central objects in our
simulations have $ M_{\rm vir} > M$, and thus are \emph{not}
gravitationally bound. This is in agreement with the observational
results of Tachihara et al.\ (2002) and Morata et al.\ (2002) for
starless cores.  However, also some observed ``starless'' cores seem
to lie close to or even fall below the equipartition line (e.g.,
Caselli et al.\ 2002), and we speculate that at least some of them may
contain as yet undetected young stellar or sub-stellar objects, as is
in the case of core L1014 (Young et al.\ 2004).

The fact that cores in the gravoturbulent model are initially created
and confined by the ram pressure from convergent larger-scale flows
leads to the velocity dispersion being highest in the low
column-density gas at the surface of the clump, either at localized
positions or with bow-like shapes. Such structure is often found in
detailed high-resolution velocity maps of observed starless cores
(e.g., Barranco \& Goodman 1998; Goodman et al.\ 1998; Caselli et al.\
2002; Tafalla et al.\ 2004). However, it is not predicted or explained
within the framework of star-formation models based on the slow
contraction of magnetically subcritical cores mediated by ambipolar
diffusion.

Our calculations support an evolutionary sequence where a molecular
cloud core is formed by turbulent ram pressure compression. As it
gains mass and becomes denser, both its internal kinetic energy and the
absolute value of its gravitational energy increase. For some cores,
the gravitational attraction may exceed any opposing forces (either
thermal or magnetic), and the core goes into collapse to quickly build
up a protostellar object in its interior. The fact that star-forming
cores are often observed near energy equipartition is only the
signature of gravity becoming dynamically important, not of
hydrostatic equilibrium. However, if the external turbulent
compression ends before the core reaches the state where it is
dominated by self-gravity, then it will reexpand and merge with the
lower-density ambient molecular cloud material.

In a typical turbulent cloud environment, the evolution of molecular
cloud cores is both transient and fast. This holds for their formation
by convergent flows, as well as for their destruction either by
collapse and transformation into stars, or by reexpansion or
dispersion by passing shock fronts.  Both, observational estimates of
pre-stellar core lifetimes in well-characterized star forming regions
(e.g., Lee \& Myers 1999; Jijina et al.\ 1999) and numerical
simulations of star-forming turbulent clouds (e.g.\ V\'azquez-Semadeni
et al.\ 2005) indicate that the process of core formation and collapse
is fast ($\lesssim 10^6\,$yr).  \footnote{Note that it is often
  thought that transient cores cannot form stars, while star-forming
  cores are the ones ``that last long enough'' to do so. For example,
  Tassis \& Mouschovias (2004) suggest that typical molecular cloud
  cores undergo a lengthy gestation period before becoming observable
  and being able to form stars.  In their picture, starless cores are
  simply not old enough yet to begin forming stars.  They thus
  conclude that observational estimates of core lifetimes do not
  constrain the duration of such lengthy early phase. However, this
  picture has problems: observationally, the existence of such a long
  gestation period would imply that most molecular clouds should
  appear devoid of star formation, in contradiction with observational
  facts (Hartmann 2003; Ballesteros-Paredes \& Hartmann 2005).
  Theoretically, the long gestation period can only occur in a
  quiescent, unperturbed medium.  The presence of supersonic
  turbulence speeds up the formation and evolution of the density
  fluctuations (i.e.\ of the cores; see Li \& Nakamura 2004;
  V\'azquez-Semadeni et al.\ 2005), independently of whether they end
  up collapsing or re-dispersing.}

We conclude that the quiescent, coherent, and sometimes near-virial
nature of observed molecular cloud cores is in direct agreement with
the theory of gravoturbulent star formation (Mac~Low \& Klessen 2004,
and references therein), where stars build up from material in the
gravitationally unstable parts of the spectrum of transient,
dynamically-evolving density fluctuations that are the characteristics
of self-gravitating, supersonically turbulent media such as
interstellar gas clouds.

\acknowledgements

We thank L.\ Hartmann, S.\ Lizano, and P., \ Myers for
stimulating discussions during the elaboration of the present
manuscript. We also thank our referee A.\ Goodman for insightful
comments and suggestions. RSK acknowledges support by the Emmy Noether
Program of the Deutsche Forschungsgemeinschaft (DFG: grant KL1358/1). JBP
and EVS acknowledge support from CONACYT's grants 27752-E and 36571-E
respectively.  This research has made use of NASA's Astrophysics Data
System Abstract Service.

\newpage

\newpage

\begin{table}[htp]
\caption{Statistics of coherent cores\label{tab:coherence}}
\begin{center}
\begin{tabular}{lrrr} 
\tableline \tableline
                    & model    & model    & complete \\  
                    & LSD      & SSD      & sample   \\  
\tableline
coherent subsonic   &51.5$\,$\%&12.9$\,$\%&24.6$\,$\%    \\
coherent transsonic &13.6$\,$\%&30.0$\,$\%&25.0$\,$\%    \\
coherent supersonic &6.0$\,$\% &13.5$\,$\%&11.3$\,$\%    \\
incoherent          &28.7$\,$\%&43.6$\,$\%&39.1$\,$\%    \\
\tableline
number of maps      & 132  & 303  & 435      \\  
\tableline
\end{tabular}
\end{center}
\end{table}

\newpage

\begin{thebibliography}{}


\bibitem[Alves et al.(2001)]{Alves_etal01} Alves, J., Lada,
C.~J., \& Lada, E.~A.\ 2001, \nat, 409, 159

\bibitem[Andre et al.(2000)]{Andre_etal00} Andr{\'e},
P., Ward-Thompson, D., \& Barsony, M.\ 2000. Protostars and Planets
IV, ed, V. Mannings, A. Boss, \& S. Russell (Tucson:Univ. Arizona
Press), 59

\bibitem[Ballesteros-Paredes (2003)]{BP03}Ballesteros-Paredes,
J. 2003. In "From Observations to Self-Consistent Modeling of the 
Interstellar Medium".\ Ed.\ M.\ Avillez \& D.\ Breitschwerdt (Kluwer),
in press.

\bibitem[Ballesteros-Paredes et al.(1999b)]{BHV99}
Ballesteros-Paredes, J., Hartmann, L., \& V{\' a}zquez-Semadeni, E.\
1999b, \apj, 527, 285

\bibitem[Ballesteros-Paredes \& Mac Low(2002)]{BPML02} 
Ballesteros-Paredes, J.~\& Mac Low, M.\ 2002, \apj, 570, 734

\bibitem[Ballesteros-Paredes et al.(1999a)]{BVS99}
Ballesteros-Paredes, J., V{\' a}zquez-Semadeni, E., \& Scalo, J.\
1999a, \apj, 515, 286

\bibitem[Barranco \& Goodman(1998)]{Barranco_Goodman98} Barranco,
J.~A.~\& Goodman, A.~A.\ 1998, \apj, 504, 207


\bibitem[\protect\citeauthoryear{Bate \&
    Burkert}{1997}]{1997MNRAS.288.1060B} Bate M.\ R., \& Burkert, A.
  1997, \mnras, 288, 1060


\bibitem[Bate, Bonnell \& Price (1995)]{Bate_etal95} Bate, M.\ R.,
Bonell, I.A.\ ., \& Price, N.\ M. 1995. \mnras, 277, 362 

\bibitem[Benz(1990)]{Benz90} Benz, W. 1990, in The Numerical Modeling
of Nonlinear Stellar Pulsations, ed. J.~R.~Buchler, p.\ 269, Kluwer
Academic Publishers, The Netherlands

\bibitem[Bonnor(1956)]{Bonnor56} Bonnor, W.~B.\ 1956, \mnras, 116, 351


\bibitem[Caselli et al.(2002)]{Caselli_etal02} 
Caselli, P., Benson, P.~J., Myers, P.~C., \& Tafalla, M.\ 2002, \apj,
572, 238


\bibitem[Crutcher(2004)]{Crutcher04} Crutcher, R.~M.\ 2004, in Magnetic
Fields and Star Formation: Theory versus Observations, eds. Ana I Gomez
de Castro et al. (Dordrecht: Kluwer), in press


\bibitem[Ebert(1955)]{Ebert55} Ebert, R.\ 1955, Zeitschrift f{\"u}r
Astrophysik, 36, 222

\bibitem[Elmegreen(1993)]{Elmegreen93} Elmegreen, B.~G.\ 1993, 
\apjl, 419, L29 


\bibitem[Gammie, Lin, Stone, \& Ostriker(2003)]{Gammieetal2003} 
Gammie, C.~F., Lin, Y., Stone, J.~M., \& Ostriker, E.~C.\ 2003, \apj, 592, 
203 

\bibitem[Goodman et al.(1998)]{Goodman_etal98} 
Goodman, A.~A., Barranco, J.~A., Wilner, D.~J., \& Heyer, M.~H.\ 1998,
\apj, 504, 223 



\bibitem[Heitsch et al.(2001)]{HMK01} Heitsch, F., Mac Low, M.-M., \& 
Klessen, R.\ S.\ 2001, \apj, 547,  280  

\bibitem[Hunter \& Fleck(1982)]{Hunter_Fleck82} Hunter, J.~H.~\& 
Fleck, R.~C.\ 1982, \apj, 256, 505

\bibitem[Jappsen \& Klessen(2004)]{JK04} Jappsen, A.-K., \& Klessen, R.\ S.\
  2004, \aap, in press (astro-ph/0402361)

\bibitem[Jappsen \& Klessen(2004)] Jappsen, A.-K., Klessen, R.\ S., 
Larson, R.,  Li, Y., \& Mac Low, M.-M. 2004. in preparation

\bibitem[Jijina, Myers, \& Adams(1999)]{Jijina_etal99} Jijina, J., 
Myers, P.~C., \& Adams, F.~C.\ 1999, \apjs, 125, 161 

\bibitem[Klessen(2001)]{Klessen01} Klessen, R.~S.\ 2001, \apj, 556,
837  

\bibitem[Klessen \& Burkert(2000)]{Klessen_Burkert00} Klessen,
R.~S.~\& Burkert, A.\ 2000, \apjs, 128, 287

\bibitem[Klessen \& Burkert(2001)]{Klessen_Burkert01} Klessen,
R.~S.~\& Burkert, A.\ 2001, \apj, 549, 386  

\bibitem[Klessen et al.(1998)]{KBB98} Klessen, R.\ S.,
Burkert, A., Bate, M.\ R. 1998, \apj, 501, L205 

\bibitem[Klessen  et al.(2000)]{KHM00} Klessen, R.~S.,
Heitsch, F., \& Mac Low, M.-M.,\ 2000, \apj, 535, 887   

\bibitem[Larson(1981)]{Larson81} Larson, R.~B.\ 1981, \mnras, 
194, 809

\bibitem[Larson(2003)]{LAR03} Larson, R.~B.\ 2003, Reports of Progress in
  Physics, 66, 1651


\bibitem[Lizano \& Shu(1989)]{Lizano_Shu89} Lizano, S.~\& Shu, 
F.~H.\ 1989, \apj, 342, 834 

\bibitem[Li et al.(2004)]{Li_etal04} Li, P.\ S., Norman, M.\ L., Mac~Low, M.-M., \&
  Heitsch, F. 2004, \apj, 605, 800

\bibitem[Mac Low \& Klessen(2004)]{RMP} Mac~Low, M.-M., \& Klessen,
  R.\ S. 2004, Rev.\ Mod.\ Phys., 76, 125

\bibitem[Mac Low \& Ossenkopf(2000)]{MacLow_Ossenkopf00} Mac Low, M.-M.~\& 
Ossenkopf, V.\ 2000, \aap, 353, 339

\bibitem[Monaghan(1992)]{Monaghan92} Monaghan, J.~J. 1992, \araa, 30,
543  

\bibitem[Morata et al.(2004)]{Morata_etal04} Morata, O., Girart,
J.~M., \& Estalella, R. 2004. \aa, submitted.


\bibitem[Myers(1983)]{Myers83} Myers, P.~C.\ 1983, \apj, 270, 
105 

\bibitem[Myers \& Goodman(1988a)]{Myers_Goodmana} Myers, P.~C.~\& 
Goodman, A.~A.\ 1988, \apjl, 326, L27 

\bibitem[Myers \& Goodman(1988b)]{Myers_Goodmanb} Myers, P.~C.~\& 
Goodman, A.~A.\ 1988, \apj, 329, 392 

\bibitem[Myers \& Fuller(1992)]{Myers_Fuller92} {Myers}, P.~C. and
{Fuller}, G.~A.. 1992. \apj, 396, 631 

\bibitem[Ossenkopf \& Mac Low(2002)]{Ossenkopf_MacLow02} Ossenkopf, V.~\& 
Mac Low, M.-M.\ 2002, \aap, 390, 307 

\bibitem[Ostriker, Stone, \& Gammie(2001)]{Ostriker_etal01} Ostriker, 
E.~C., Stone, J.~M., \& Gammie, C.~F.\ 2001, \apj, 546, 980 

\bibitem[Padoan(1995)]{Padoan95} Padoan, P.\ 1995, \mnras, 277, 
377 

\bibitem[Padoan et al.(2001)]{Padoan_etal01b} Padoan, P., Goodman, A., 
Draine, B.~T., Juvela, M., Nordlund, {\AA}., \& R{\" o}gnvaldsson, {\" 
O}.~E.\ 2001b, \apj, 559, 100

\bibitem[Padoan et al.(2001)]{Padoan_etal01a} 
Padoan, P., Juvela, M., Goodman, A.~A., \& Nordlund, {\AA}.\ 2001a,
\apj, 553, 227


\bibitem[Padoan \& Nordlund(2002)]{Padoan_Nordlund02} Padoan, P.~\& 
Nordlund, {\AA}.\ 2002, \apj, 576, 870 

\bibitem[Reach et al.(2004)]{2004ApJS..154..385R} Reach, W.~T., et al.\ 
2004, \apjs, 154, 385

\bibitem[Sasao (1973)]{Sasao73} Sasao, T. 1973. PASJ, 25, 1

\bibitem[Schmeja \& Klessen(2004)]{SK04} Schmeja, S., \& Klessen, R.\ S.\
  2004, \aap, 419, 405

\bibitem[Shu et al.(1987)]{SAL} Shu, F.~H.,  Adams, F.~C., \& Lizano, S.\ 1987,
\araa, 25, 23  

\bibitem[Swade(1989)]{Swade89} Swade, D.~A.\ 1989, \apj, 345, 
828

\bibitem[Tachihara et al.(2002)]{Tachihara02} Tachihara, K., Onishi, T.,
  Mizuno, A., \& Fukui, Y.\ 2002, \aap, 385, 909

\bibitem[Tafalla et al.\ (2004)]{Tafalla_etal04} {Tafalla}, M. and
{Myers}, P.~C. and {Caselli}, P. and {Walmsley}, C.~M. 2004. {\aap},
416, 191


\bibitem[Tilley \& Pudritz(2004)]{TP04} Tilley, D.\ A.\& Pudritz, R.\ 
  E. 2004, \mnras, in press (astro-ph/0406122)

\bibitem[V\'azquez-Semadeni et al.\ 2000]{VS_etal00} V\'azquez-Semadeni,
E., Ostriker, E. C., Passot, T., Gammie, C. \& Stone, J. 2000, in
Protostars \& Planets IV, ed. V. Mannings, 
A. Boss \& S. Russell (Tucson: Univ.\ of Arizona Press), 3

\bibitem[V\'azquez-Semadeni, Ballesteros-Paredes \& Klessen(2003a)]
{VBK03a} V{\' a}zquez-Semadeni, E., 
Ballesteros-Paredes, J., \& Klessen, R.~S.\ 2003a, \apjl, 585, L131  


\bibitem[V\'azquez-Semadeni et al.(2004)]
{VKSB04} V\'azquez-Semadeni, E., Kim, J., Shadmehri, M., \&
Ballesteros-Paredes, J. 2004, \apj, in press


\bibitem[von Weizs{\"a}cker (1951)]
{Weiszacker51} von Weizs{\"a}cker, C.F. 1951. \apj,  114, 165

\bibitem[\protect\citeauthoryear{Wuchterl \&
    Klessen}{2001}]{2001ApJ...560L.185W} Wuchterl, G., \& Klessen, R.\ 
  S. 2001, \apj, 560, L185


\bibitem[{{Young} {et~al.}(2004){Young}, {Joergensen}, {Shirley}, {Kauffmann},
  {Huard}, {Lai}, {Lee}, {Crapsi}, {Bourke}, {Dullemond}, {Brooke},
  {Porras}, {Spiesman}, {Allen}, {Blake}, {Evans II}, {Harvey},
  {Koerner}, {Mundy}, {Myers}, {Padgett}, {Sargent}, {Stapelfeldt},
  {van Dishoeck}, {Bertoldi}, {Chapman}, {Cieza}, {DeVries}, {Ridge},
  \& {Wahhaj}}]{2004astro.ph..6371Y} {Young}, C.~H., {Joergensen},
  J.~K., {Shirley}, Y.~L., {Kauffmann}, J., {Huard}, T., {Lai}, S.~.,
  {Lee}, C.~W., {Crapsi}, A., {Bourke}, T.~L., {Dullemond}, C.~P.,
  {Brooke}, T.~Y., {Porras}, A., {Spiesman}, W., {Allen}, L.~E.,
  {Blake}, G.~A., {Evans II}, N.~J., {Harvey}, P.~M., {Koerner},
  D.~W., {Mundy}, L.~G., {Myers}, P.~C., {Padgett}, D.~L., {Sargent},
  A.~I., {Stapelfeldt}, K.~R., {van Dishoeck}, E.~F., {Bertoldi}, F.,
  {Chapman}, N., {Cieza}, L., {DeVries}, C.~H., {Ridge}, N.~A., \&
  {Wahhaj}, Z. 2004. ApJL, in press

\end{thebibliography}
\end{document}